# Efficient and Accurate Tree Detection from 3D Point Clouds through Paid Crowdsourcing

Michael Kölle, Volker Walter, Ivan Shiller, Uwe Soergel

Institute for Photogrammetry, University of Stuttgart, Germany

email: {firstname.lastname}@ifp.uni-stuttgart.de

Corresponding author: volker.walter@ifp.uni-stuttgart.de

## Abstract

Accurate tree detection is of growing importance in applications such as urban planning, forest inventory, and environmental monitoring. In this article, we present an approach to creating tree maps by annotating them in 3D point clouds. Point cloud representations allow the precise identification of tree positions, particularly stem locations, and their heights. Our method leverages human computational power through paid crowdsourcing, employing a web tool designed to enable even non-experts to effectively tackle the task. The primary focus of this paper is to discuss the web tool's development and strategies to ensure high-quality tree annotations despite encountering noise in the crowdsourced data. Following our methodology, we achieve quality measures surpassing 90% for various challenging test sets of diverse complexities. We emphasize that our tree map creation process, including initial point cloud collection, can be completed within 1-2 days.

## Acknowledgment

This work profited from funding through the German Research Foundation as part of Germany´s Excellence Strategy – EXC 2120/1 – 390831618.

# 1 Introduction

While manual tree detection is time-consuming and expensive, automated methods are limited by the complexity of the environment, the variability of tree shapes, and the quality of the data. To overcome these challenges, we propose a method that relies on human computation power in terms of paid crowdsourcing. To overcome the inherent data quality inhomogeneity of this technique, we exploit the Wisdom of the Crowds principle, which translates to capturing each tree multiple times by multiple crowdworkers, allowing automatic outlier detection and a significant quality improvement of the results compared to individual acquisitions.

The annotation of trees from 3D point clouds has already been subject of our previous work (Walter et al., 2020). So far, however, we have only applied our method to a few exemplary, manually selected datasets to demonstrate the feasibility in principle. Hence, we see this paper as natural consequence with special emphasis on applying our (adapted) procedures to various data sets depicting scenes of different complexity. Now that we are faced with non-handpicked complex scenes, we have improved our web tool in several ways. This web tool is one of the most crucial components when outsourcing tasks to non-expert annotators as it needs to be as easy and intuitive to handle as possible while still allowing sufficient interaction with 3D data. In this regard, enhancements mainly focus on adapting to extended point clouds, often with a high density of closely standing (intertwined) trees. Mainly, this can be solved by generating profile-shaped tiles from point cloud data, which are then sent to the crowd for labelling, allowing us to achieve satisfactory results at a quality level comparable to that of dedicated experts.

The rest of this paper is organized as follows. After reviewing related work with respect to crowdsourced data acquisition and possible implementations of the Wisdom of the Crowds principle (Section 2), Section 3 will outline our methodology with special emphasis on data pre-processing, implementation of the graphical user interface and data integration routines. Applying those techniques to various data sets presented in Section 4, we are able to generate competitive results discussed in Section 5. Finally, Section 6 will summarize our findings and discuss potential future research directions.

# 2 Related Work

Crowdsourcing is an artificial term created by merging the words "crowd" and "outsourcing", introduced by Howe (2006). While outsourcing means transferring specific tasks to specified individuals, in crowdsourcing typically tasks are delegated to anonymous workers, commonly referred to as crowdworkers, through online platforms. This approach provides employers with access to a vast pool of workers who might otherwise remain inaccessible.

Many prominent crowdsourcing projects rely on the participation of unpaid volunteers, such as Wikipedia (www.wikipedia.org) and Zooniverse (www.zooniverse.org). The voluntary collection of geographic data is known as Volunteered Geographic Information (VGI) (Goodchild, 2007). OpenStreetMap (OSM - www.openstreetmap.org) is a well-known VGI project that aims to create a map of the world to which anyone can contribute (Haklay and Weber, 2008). Successful VGI projects require an active and motivated community of participants. Budhathoki and Haythornthwaite (2012) provide an extensive discussion of motivational factors for participation in VGI projects.

When such intrinsic motivations are not present, we need to rely on extrinsic incentives to encourage participation. A common method for this is to offer financial compensation (Haralabopoulos et al., 2019). In paid crowdsourcing, tasks are most conveniently handled through online marketplaces such as microWorkers (www.microworkers.com) (Hirth et al., 2011) or Amazon Mechanical Turk (MTurk - www.mturk.com) (Ipeirotis, 2010), responsible for the recruitment and payment of the workers. Possible applications of paid crowdsourcing range from image-based polyp detection (Nguyen et al., 2012), image aesthetics evaluation (Redi and Povoa, 2014), translation tasks (Gao et al., 2015) to content creation for museum apps (Lans et al., 2018) and traffic monitoring (Koita and Suzuki, 2019).

However, when employing annotators with different skills, education and background knowledge, often, we are confronted with a high level of data inhomogeneity (Vaughan, 2017; Daniel et al., 2018). This is especially true for paid crowdsourcing, where there may be dishonest workers trying to maximize their income by submitting as many tasks as possible, but providing incomplete or poor results (Hirth et al., 2011). Getting accurate and high-quality results from non-experts is a major challenge in crowdsourcing (Zhou et al. 2012; Leibovici et al., 2017; Liu et al., 2018).

There are two options for improving the quality of crowdsourced data (Zhang et al., 2016): i) Quality Control on Task Designing and ii) Quality Improvement after Data Collection. For the first approach, techniques such as qualification tasks, control tasks and tutorials are commonplace. A detailed overview of such methods can be found in the work of Daniel et al. (2018). Although these methods are likely to improve the overall quality of crowdsourcing campaigns, there will still be results that are exceptionally good or bad. This is especially the case for geospatial data acquisition as crowdworkers are typically not familiar with the standards of geospatial data collection (Hashemi and Abbaspous, 2015) and often, even dedicated experts disagree on the true annotation (Walter and Soergel, 2018).

As for the second option, Quality Improvement after Data Collection, respective methods often rely on collecting data multiple times by different crowdworkers in order to infer the true result and eliminate outliers in the process, referred to as "truth inference" (Zhen et al., 2017). Surowiecki (2005) demonstrates for many different applications that aggregating results from multiple contributors can lead to a result that even exceeds the best individual (expert) contribution. This approach is known as "The Wisdom of the Crowds". However, this requires solving tasks multiple times (thus also increasing costs when dealing with paid crowdsourcing) and formulating a specific aggregation rule (Simons, 2004).

The simplest realization of the Wisdom of the Crowds principle is majority voting (Juni and Eckstein, 2017), which is based on the assumption that the majority of workers are trustworthy (Kazemi et al., 2013). In this approach, the employer assigns the same task to multiple crowdworkers, assuming that the true answer aligns with the majority vote (Hirth et al., 2013). Majority voting is a simple but effective technique (Zhang et al., 2016) and has been used in various spatial labelling tasks. For example, cropland identification in remote sensing images (Salk et al., 2016), classification of building footprint data (Hecht et al., 2018), accessibility map generation (Liu et al., 2018) and crowd-based labelling of 3D LiDAR point clouds (Koelle et al., 2020). For localization tasks, such as tree annotation, we cannot perform simple majority voting, but compute an average position from the crowdsourced data, e.g., based on a DBSCAN clustering procedure (Walter et al., 2020).

For the aforementioned adaptions of our previous methodology (Walter et al., 2020), we were influenced by the work of Herfort et al. (2018), who seek tree annotations from crowdworkers by means of providing them with small point cloud subsets with a point-cloud-section-like appearance. Precisely, the authors employ crowdworkers for the classification of such subsets into a set of categories and conduct crown base estimation by means of averaging answers from multiple crowdworkers.

# 3 Methodology

## 3.1 Data Acquisition & Pre-Processing

Within this work, we intend to generate tree cadastre data by developing a method that is i) independent of any additional data and ii) that runs in a manner where expert involvement can be reduced to a minimum. To achieve a maximum level of independence, we propose capturing point clouds with a mobile laser scanning system like the *GeoSlam Zeb Horizon* scanner (although data from other sources, such as airborne platforms, can be processed in the same way), which realizes a SLAM-based acquisition built upon a *Velodyne Puck VLP-16* laser scanner (https://geoslam.com/solutions/zeb-horizon/). The main advantage using this system is the flexibility to acquire any scene of interest in a matter of minutes (all test sites described in Section 4 were captured in less than 30 minutes). Hence, the methodology presented in the following can be easily applied to other sites.

As a next step, the captured point cloud data is to be prepared for the web tool (cf. Section 3.2) and the crowd, respectively. In this regard, we refrain from picking representative circular subsets with 50 m radii like proposed in our previous work (Walter et al., 2020), and modify the generation of tiles to be suitable for large-scale acquisition campaigns. This is accomplished i) by utilizing rectangular tiles and ii) by shaping them to be more similar to profile sections. While the first aspect mainly allows covering extended areas more efficiently, the latter is key to the endeavour of crowd-based tree acquisition. In densely vegetated regions, crowdworkers would be simply overwhelmed with the amount of trees in wide areas (cf. Figure 1 (a)), but even experts would fail to succeed due to occlusions of individual trees. In this regard, reducing the spatial extent of tiles can already help, but occlusions will nevertheless remain (cf. Figure 1 (b)). Consequently, we reduce the tile size even further, and turn from quadratic shapes to point cloud profiles (cf. Figure 1 (c)) where individual stems become clearly visible.

Apart from extracting such subsets, the raw point cloud data is subsampled to 20 cm point spacing to minimize the data footprint and to allow for quick loading of data by crowdworkers utilizing the web tool. Furthermore, as visible from Figure 1, we create a pseudo-colorization with a height-above-ground colour scheme to allow for easier interpretation. This requires having either a Digital Terrain Model (DTM) of the area of interest, which can be obtained from national mapping agencies, or alternatively, can be generated on the fly with respective software packages such as SCOP++ (Pfeifer et al., 2001).

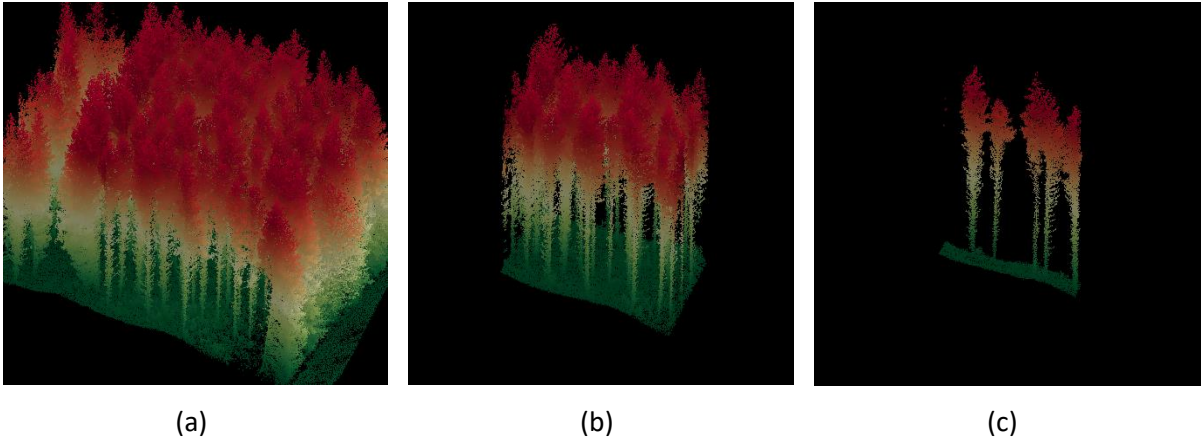

Figure 1: Exemplary subsets from one of the densely vegetated areas with a tile size of (a) 50m x 50m, (b) 25m x 25m and (c) 25m x 4m.

With respect to the type of annotations desired, we advocate marking tree stems and their height using cylinders (with a fixed radius of 0.5 m). We would like to emphasize that we refrain from asking crowdworkers to record the exact radii of tree crowns (e.g., by using minimal bounding cylinders as in the work of Walter et al. (2020), as this is a less clear task that allows for a wide range of possible valid solutions (e.g., try to imagine vertical bounding lines that clearly separate the tree crowns in Figure 1 (c)).

## 3. 2 Graphical User Interface

If an employer aims to employ crowdsourcing, easy-to-use tools are to be provided. This is especially the case if crowdworkers are to work with probably very unfamiliar data such as 3D point clouds. Therefore, we have improved an initial prototype (Walter et al., 2020) for this purpose in several ways, as described below. In addition, we recommend interested readers to try a demo version of our tool, which is available at https://crowd.ifp.uni-stuttgart.de/DEMO_Trees/main_local.php.

As motivated in the previous section, we are interested in annotating trees by means of cylinders (with fixed radii) at the stem location and adjusting the height of the cylinder to fit the stem. Hence, the basic working principle of our web tool also follows this scheme. Precisely, we distinguish between a control panel (Figure 2, right) and three data panels (Figure 2, left). Working with our tool, a crowdworker would first create a new cylinder (spawning in the point cloud centre with default height), then use the data panels to precisely position the cylinder at the correct location and afterwards scale the height by usage of the respective buttons in the control panel.

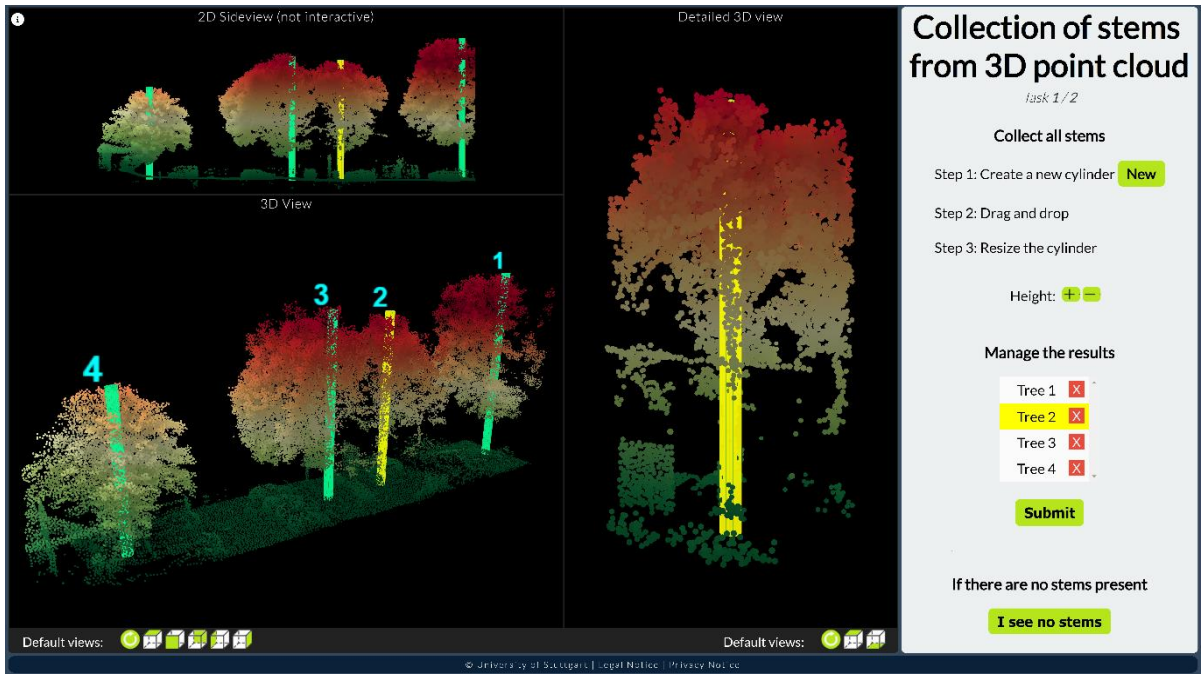

Figure 2: Our web-based graphical user interface for the acquisition of trees from 3D point clouds. The cylinder that is currently under review is highlighted in yellow and a 5 m cylindrical neighbourhood is cropped and visualized in detail in the right panel. Cylinders that have already been set are indicated in *green*, but can be altered at any time afterwards.

For the positioning of the cylinders, we provide three different data panels. This includes an interactive 3D view (cf. Figure 2, *bottom left*) that allows to freely interact and explore the point cloud and is intended to be used for an initial coarse positioning, i.e., crowdworkers are asked to move the cylinder from the initial position to a stem location by means of a simple drag and drop operation. Cylinder bottoms are automatically set to the local DTM height (known from providing respective DTM tiles, cf. Section 3.1). The second data panel shows a static 2D side view that can be interpreted as a projection of the point cloud data to the *xz*-plane (cf. Figure 2, *top left*). This data panel gives an overview of the complete data set as well as a sense of how many trees are included and to be marked.

The next step is to refine the already (roughly) set cylinder, which can be achieved either by adjusting the cylinder position in the 3D view panel or by using the third data panel, which presents a detailed 3D view (cf. Figure 2, *centre*). We extract a subset of the point cloud around 5 m of the current stem and hide all points near the ground by masking points that are among the 5% lowest points in this point cloud subsets. This allows a bottom-up view in which stems are easily recognizable.

As we might encounter tiles that do not include any trees at all we give crowdworkers the option to mark tiles as including no trees (cf. Figure 2, "I see no stems"-button), which leads to presenting the crowdworker alternative point clouds until annotations are made. This way, we can minimize the risk of workers misusing this option to quickly skip through the task and being paid undeservedly.

Aside from the methods described above, for generating accurate data, quality control methods are critical in paid crowdsourcing. As mentioned in Section 2, we distinguish between i) Quality Control on Task Designing and ii) Quality Improvement after Data Collection.

With regard to the first option, we have developed a tool suitable for achieving high-quality results and train crowdworkers in the proper use of this tool by means of short introductory videos. Additionally, each of the crowd jobs consists of two point clouds that are to be annotated. For the first one, Ground Truth (GT) data is available and is utilized to assess the quality of acquired data. Only if the acquired data match the GT data in terms of the correct number of stems, a maximum position discrepancy of 1 m and a maximum height discrepancy of 2 m, we accept the contribution as valid submission and pay crowdworkers. The subsequent second point cloud is the actual payload job, for which no GT is available.

To improve quality after data collection, we present each of the tiles to multiple crowdworkers and then integrate the results as described in the following section.

## 3.3 Integration of Results in Post-Processing

Since our crowdworkers capture the point cloud multiple times, we need to find a way to integrate these multiple acquisitions while eliminating outliers. Here, we follow the method of Walter et al. (2020), but adapt it to a stem-based acquisition scheme.

The basic idea is to merge all crowd acquisitions and identify clusters, which are supposed to form around true tree stem positions. Since we are unaware of the true number of trees in the data set, we utilize the DBSCAN algorithm (Ester et al., 1996) for clustering, for which this knowledge is not required. DBSCAN, on the other hand, is parametrized by the maximum distance $epsilon$ that is considered for neighbour identification and the minimum number of instances $n_{min}$ that are required to form a cluster. In a nutshell, DBSCAN visits all points in a data set and classifies them either as noise (if $< n_{min}$ points are within distance $epsilon$), core points (if $\geq n_{min}$ are within distance $epsilon$) or border points (if at least one core point is within distance $epsilon$), with the latter two actually contributing to clusters. To allow an intuitive understanding of those parameters, we apply this clustering to *xy*-positions of stems only (and do not include height values at the same time). Thus, *epsilon* can be understood as minimum distance between individual clusters in the result (precisely, the minimum distance between the two closest points from the clusters) and $n_{min}$ is to be set in consideration with the number of multiple acquisitions per tile.

After the clustering step, we compute the mean position of all stems included in this cluster. However, dealing with large-scale data and various scenes, typically it is hard to find a value for *epsilon* that fits each situation depicted in the data sets. Thus, we recommend refining all clusters that are comprised by more stem acquisitions than the number of multiple acquisitions, since this indicates a value for *epsilon* that is set too large for the specific region resulting in merging stems from actually two different trees that are in close proximity to each other. Hence, if the number of acquisitions exceeds a certain threshold $n_{max}$, we initialize a second DBSCAN clustering sequence operating on only those points belonging to the respective cluster and iteratively decrease *epsilon* by 0.5 m until less than $n_{max}$ points are included in all clusters formed.

Finally, as a last step of refinement, we eliminate outliers within each cluster by comparing the cluster's median tree parameters (both position and height) to all acquisitions in this cluster and iteratively eliminate acquisitions with a position and height difference that is greater $d_{pos}$ and $d_h$, respectively. This step is repeated until all acquisitions are within these error margins compared to the median parameters.

## 4 Data

In order to prove the effectiveness of the proposed methodology in various scenarios, we collected a set of four different point clouds with different characteristics each, which are described in more detail in the following and visualized in Figure 3:

- Scene I: This point cloud represents an urban area with well-maintained tree alleys being comprised of regularly planted trees and some single standing trees.
- Scene II: Increasing the complexity, this scene depicts a cared-for city park area with trees being positioned a bit more irregularly, partly constituting rather densely vegetated areas.
- Scene III: This data set includes an old cemetery with trees being irregular in shape and height and positioned in an equally irregular pattern.
- Scene IV: This last scene is expected to be most complex, as it incorporates a densely vegetated forest scene with tree positions following a highly irregular pattern.

Each of those test sites is subject to pre-processing as discussed in Section 3.1. For tiling, the desired strip length of the point cloud profile sections is 60 m, but we allow ourselves to refrain from this value in favour of obtaining consistent, equally sized tiles for the whole data set (with close-to-desired dimensions), avoiding small remnant tiles at data borders. However, for the densely vegetated forest area (Scene IV), we created shorter and less deep sections pursuing the idea of having roughly an equal number of trees per crowd job and to avoid occlusions as motivated in Section 3.1. A detailed statistics of the data set-up including the number of GT stems can be found in Table 1. GT data was created carefully by the authors using our web tool presented in Section 3.2.

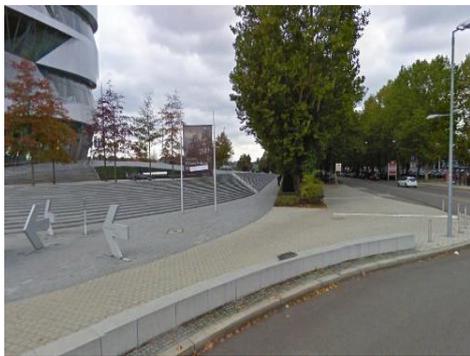 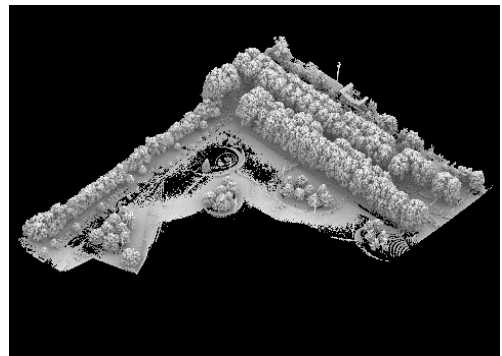

a) Scene I: Urban tree alley area

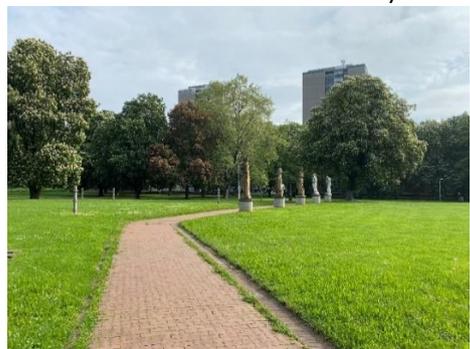 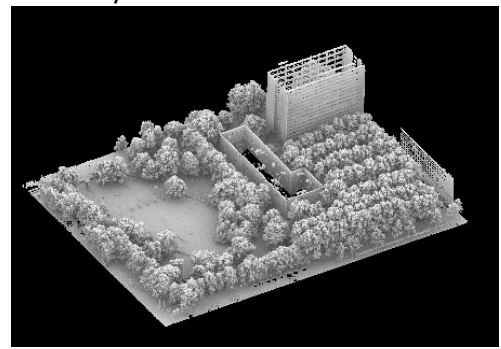

b) Scene II: Maintained city park area

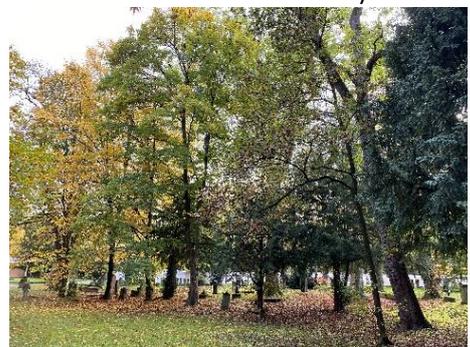 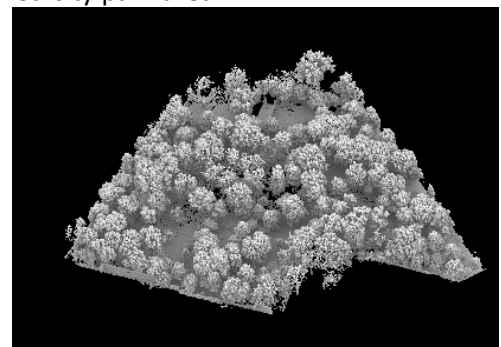

c) Scene III: Natural city park area

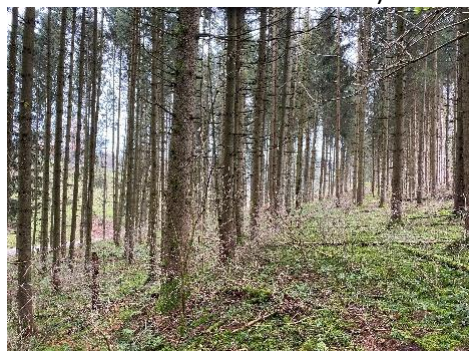 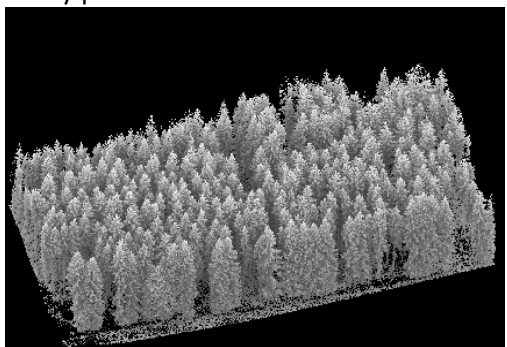

d) Scene IV: Forest area

Figure 3. An overview of the different test sites with different characteristics each. While the left column gives a visual impression of the respective data set, the right column depicts the respective area as captured as point cloud data.

**Table 1**: Data set-up for the test sites.

| Data Set | Area Size [ha] | # Tiles | Tile Size | GT Trees |
|---|---|---|---|---|
| Scene I: Urban tree alley area | 2.80 | 51 | 64 x 10 m | 146 |
| Scene II: Maintained city park area | 4.27 | 76 | 60 x 10 m | 193 |
| Scene III: Natural city park area | 2.53 | 61 | 52 x 10 m | 217 |
| Scene IV: Forest area | 0.75 | 101 | 20 x 4 m | 386 |

To get a feeling for the complexity of the different scenes, Figure 4 visualizes exemplary point cloud sections. Scene I and Scene II have similar levels of complexity, although they originate from point cloud datasets of different characteristics. Scene III seems to be more demanding, which is mainly due to a high level of beneath-crown vegetation and due to neighbouring trees overlapping/intertwining.

For the forest test site (Scene IV), on the other hand, the crowd jobs become rather straightforward as trees in deep forest are typically characterized by an extended crown and scarcely any boughs beneath. Thus, such scenes pose close-to-optimal conditions for creating easily comprehensible profile point cloud sections for a clear identification of individual tree stems. However, individual stem positions are rather close to each other, what might overwhelm crowdworkers, as stems to be set would also be in close proximity to each other. Hence, this might lead to crowdworkers missing some trees and also to a feeling of being stuck as it feels even more tedious to annotate objects with a small distance in between. As a remedy, we stretch the *xy*-coordinates by a factor of 1.5 (cf. Figure 4 (d) + (e)).

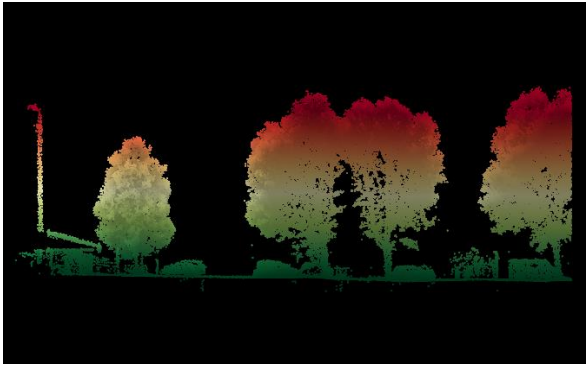
a) Scene I: Urban tree alley area

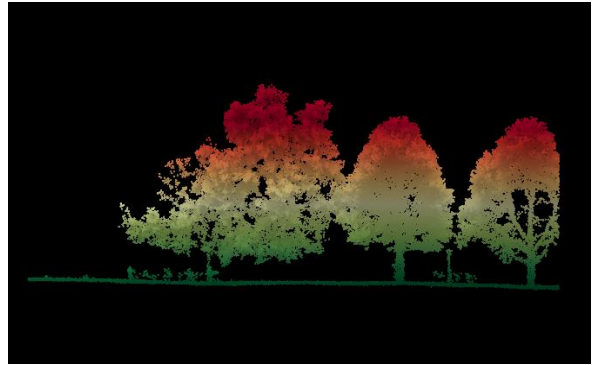
b) Scene II: Maintained city park area

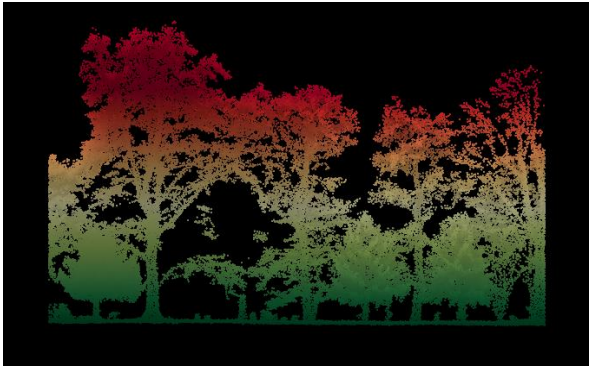
c) Scene III: Natural city park area

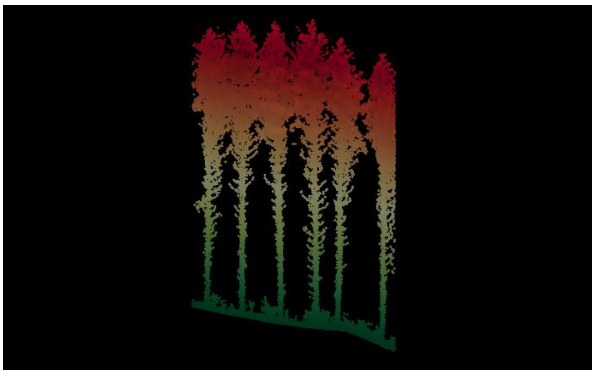
d) Scene IV: Forest area

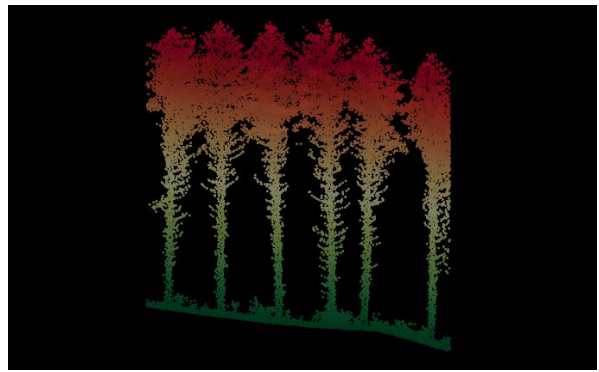
e) Scene IV: Forest area - stretched

Figure 4. Exemplary profile sections of the different test sites.

# 5 Results

For all four test sites described in the previous section, we launched a crowd campaign on the *microWorkers* platform. We offered the crowd jobs to the *Top Performers* group only. For each successfully submitted crowd job (i.e., passing the control task, cf. Section 3.2), we granted a payment of $0.10. Each job was posted to 10 different crowdworkers leading to overall campaign costs of $51/$76/$61/$101 for Scenes I-IV plus a 10% *microWorkers* fee each.

For the integration of the multiple acquisitions, we employed the integration scheme, discussed in Section 3.3, where parameters of the initial DBSCAN clustering are set to $epsilon = 1$ m in order keep acquisitions of trees in close proximity apart and $n_{min} = 4$, i.e., we expect that at least 4 out of 10 crowdworkers are capable to detect the trees. For the successive DBSCAN refinement step (cf. Section 3.3), we chose $n_{max} = 15$, i.e., assume that 15 acquisitions (and more) per cluster are a strong indication of having merged annotations from two different trees as the maximum number of acquisitions should not exceed the number of multiple acquisitions (10). To only keep high-quality acquisitions per cluster, we filter clusters as described in Section 3.3, with $d_{pos} = 1$ m and $d_h = 2$ m, i.e., we eliminate acquisitions in each cluster until none of the acquisitions has a difference to the cluster's median tree $> d_{pos}$ and $> d_h$. Please note that the same set of parameters is used for all test sites.

Figure 5 visualizes this integration process for an exemplary subset of Scene IV. As can be seen from Figure 5 (a), in most cases, crowd stems concentrate on certain spots with only a narrow scattering range, that is, whenever trees are easily distinguishable. Hence, this scattering range and related to that the spatial extent of clusters formed within this initial DBSCAN clustering step can serve as a measure for the complexity of the scene (cf. Figure 5 (b)). Apart from that, the initial clustering acts as filter for coarse outliers (cf. Figure 5 (a)). Such extended clusters also underline the importance of the second DBSCAN clustering stage to refine the segments initially set into distinct clusters while only operating on stems of each specific cluster from the first stage. As can be observed from Figure 5 (d), this allows to further purify clusters, i.e., detect outliers (cf. Figure 5 (c)) and leads to the aspired subdivision of extended clusters. Please note that an alternative single DBSCAN clustering step with a smaller $epsilon$ poses the threat of ending up with an excessive number of integrated stem positions and has proven inefficient.

Finally, the last step of purging already set clusters from crowd acquisitions that do not match the current median tree is a strict filtering stage (cf. Figure 5 (e)), necessary to determine a set of stems with a quality sufficient to contribute to the final stem derivation of each cluster (cf. Figure 5 (f)). As depicted in Figure 5 (f) and 5 (g), in most cases we succeed to compute stem positions in an accurate and complete manner. However, the integration fails, whenever stem acquisitions actually belonging to different trees are too close to each other not allowing for a clear distinction between individual acquisitions, thus ending up in integrating acquisitions of multiple trees in one stem position only (e.g., central blue cluster in Figure 5, right column).

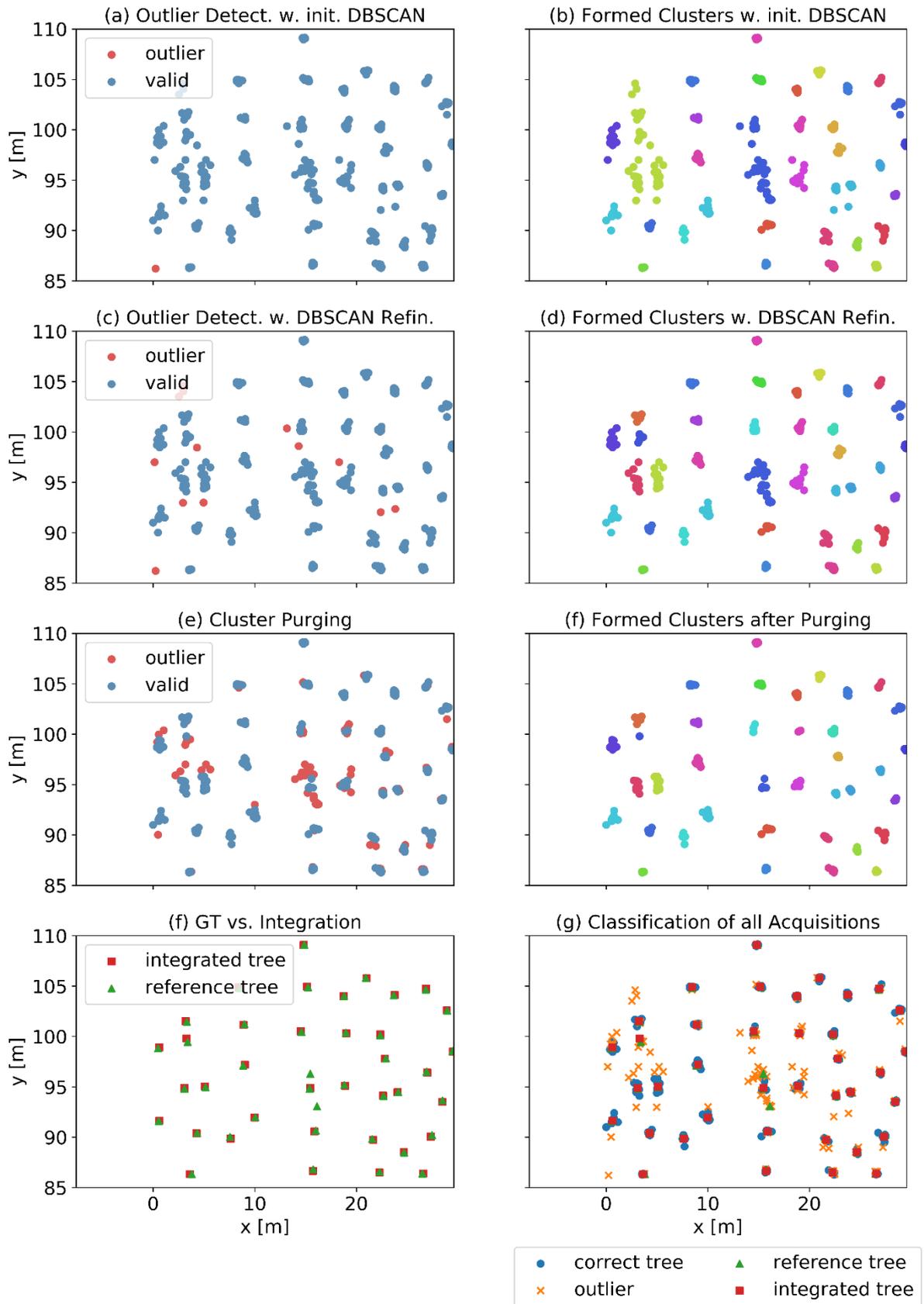

Figure 5: Step-by-step visualization of the integration process to eliminate outliers and to derive integrated stem positions.

To not only give a visual impression of the achieved results but also to assess the results quantitatively, we computed the completeness and correctness. However, this requires matching the GT stems with the integrated crowd stems. Eventually, this comes down to determining the number of True Positives (TPs), False Positives (FPs) and False Negatives (FNs). That means that we have to identify which crowd stems and GT stems, respectively, belong to which of the aforementioned categories, which is a non-trivial task as we encounter also many-to-many relationships (Walter et al., 2020).

As we are just dealing with stem positions and heights, we are limited to only comparing these parameters. This means, whenever we are confronted with a GT stem for which there is an integrated crowd acquisition with a position and height difference of $d_{pos} \leq 1$ m and $d_h \leq 2$ m (same thresholds as for the refinement process in Section 3.3), we consider it as TP. Otherwise, we are confronted with a FN. Similarly, all integrated crowd acquisitions that do not match any GT trees within these limits are marked as FPs. Based on the number of representatives for each of those categories, we finally compute Recall, Precision and Quality values as outlined in the work of Walter et al. (2020).

Table 2: Quality evaluation of the crowd-based tree mapping for the four different test sites along with the price per TP and per hectare in the paid crowdsourcing scenario.

| Data Set | GT | TP | FN | FP | Recall/ PA [%] | Precision/ UA [%] | Quality [%] | Price per TP [$] | Price per ha [$] |
|---|---|---|---|---|---|---|---|---|---|
| Scene I: Urban tree alley area | 146 | 136 | 10 | 9 | 93.15 | 93.79 | 93.47 | 0.38 | 18.21 |
| Scene II: Maintained city park area | 193 | 187 | 6 | 10 | 96.89 | 94.92 | 95.90 | 0.41 | 17.81 |
| Scene III: Natural city park area | 217 | 185 | 32 | 6 | 85.25 | 96.86 | 90.69 | 0.33 | 24.16 |
| Scene IV: Forest area | 386 | 375 | 13 | 7 | 96.63 | 98.16 | 97.39 | 0.27 | 134.90 |

From Table 2 we can observe that we can reach a quality value well over 90% for all test sites underlining the robustness and efficiency of our approach. As already assumed, Scene I and II perform rather similar, as both test sites are equally complex (cf. Figure 4 (a) & (b)). For Scene III, however, crowdworkers perform worst as this area is most demanding for identifying individual stems. This is due to crowdworkers missing a significant number of trees (low recall value). Figure 6 gives an idea with regard to which scenes are causing this effect. Obviously, such FNs occur whenever trees are in close proximity to each other (cf. Figure 6, top row). If crowdworkers do not carefully set cylinders to tree stems (but with certain offsets), this will cause regions with a high concentration of acquisitions in which the identification of individual clusters is hard to accomplish. Therefore, our integration mechanism also contributes to this effect in a way, since we will fail to distinct individual sub-clusters and we will unintentionally eliminate TP acquisitions of individual crowdworkers in the purification step (cf. Figure 5 (e)).

Secondly, FNs also occur in regions where crowdworkers are confronted with a high variation of both tree shape and tree height (cf. Figure 6, bottom row). In such cases, crowdworkers typically successfully annotate most prominent trees but miss smaller trees in between. This is understandable, as often such smaller trees are more reminiscent of beneath-crown vegetation such as shrubs. Although, we exemplify these two causes for FNs, i.e., trees in close proximity to each other and a high irregularity with regard to tree types and sizes for Scene III, similar examples also occur in the other data sets.

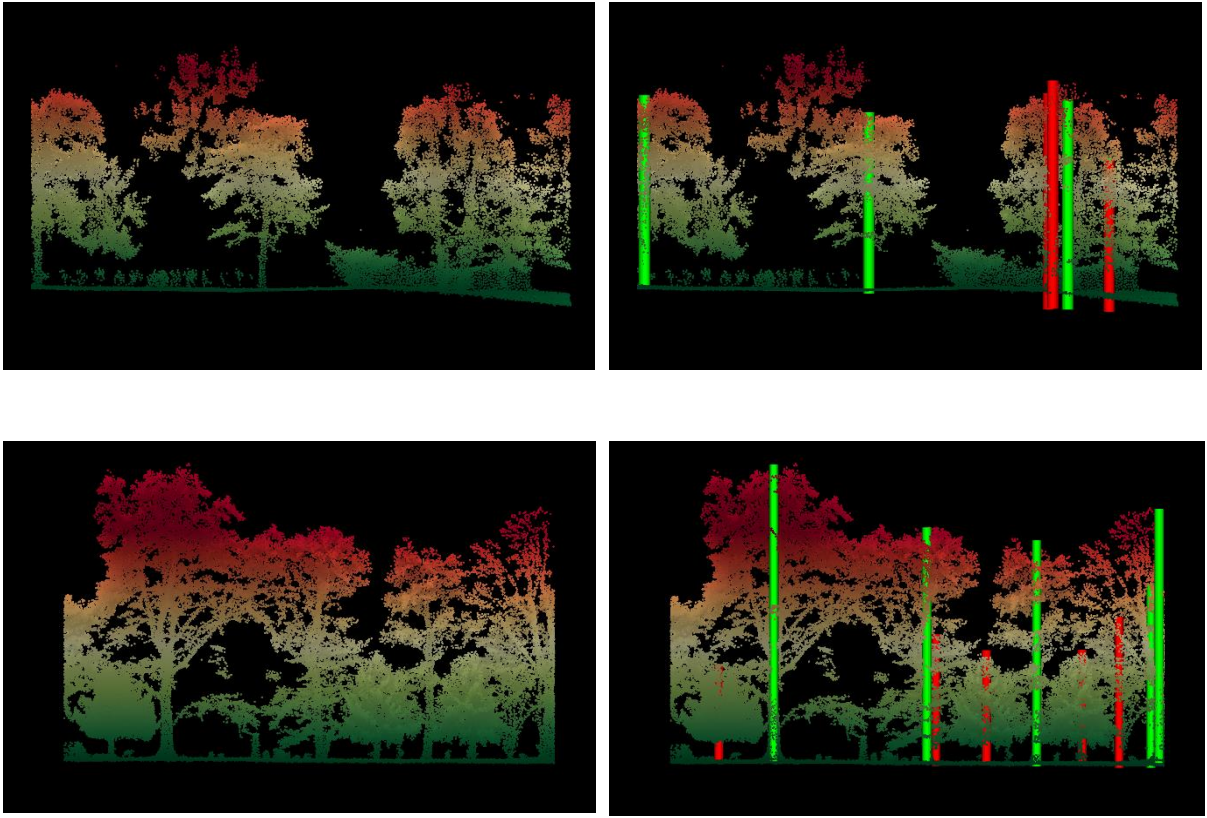

Figure 6: Exemplary point cloud profiles (left column) along with the integrated crowd acquisitions (right) classified as TP (green) and FN (red) in Scene III, which either root from trees being in close proximity to each other (top) or from a vast range of different tree sizes and shapes (bottom).

Furthermore, Figure 7 visualizes regions in which FPs occur, which, generally speaking, are scarce however. As crowdworkers are sensitive to annotating all kind of objects resembling tree stems, it is only understandable that lamp posts or wire posts will be annotated as well. However, this only holds true if these posts incorporate features that are similar to branches or tree crowns. For instance, the street light post in Figure 7 (top row) incorporates multiple light sources at different height levels. Hence, from a certain level of abstraction, we could be very well dealing with an extremely pruned tree stem.

Even harder to separate from actual tree stems are lamp posts situated in close vicinity to trees as depicted in Figure 7 (bottom row). From the 2D side view, such objects cannot be separated from actual stems and even interacting with such regions in the 3D views requires a high level of point cloud familiarity and experience. Nevertheless, for such complex scenes also experts would most likely achieve different outcomes. However, we would like to stress that other lamp posts like the ones on the left in Figure 7 (bottom row) correctly lack an annotation.

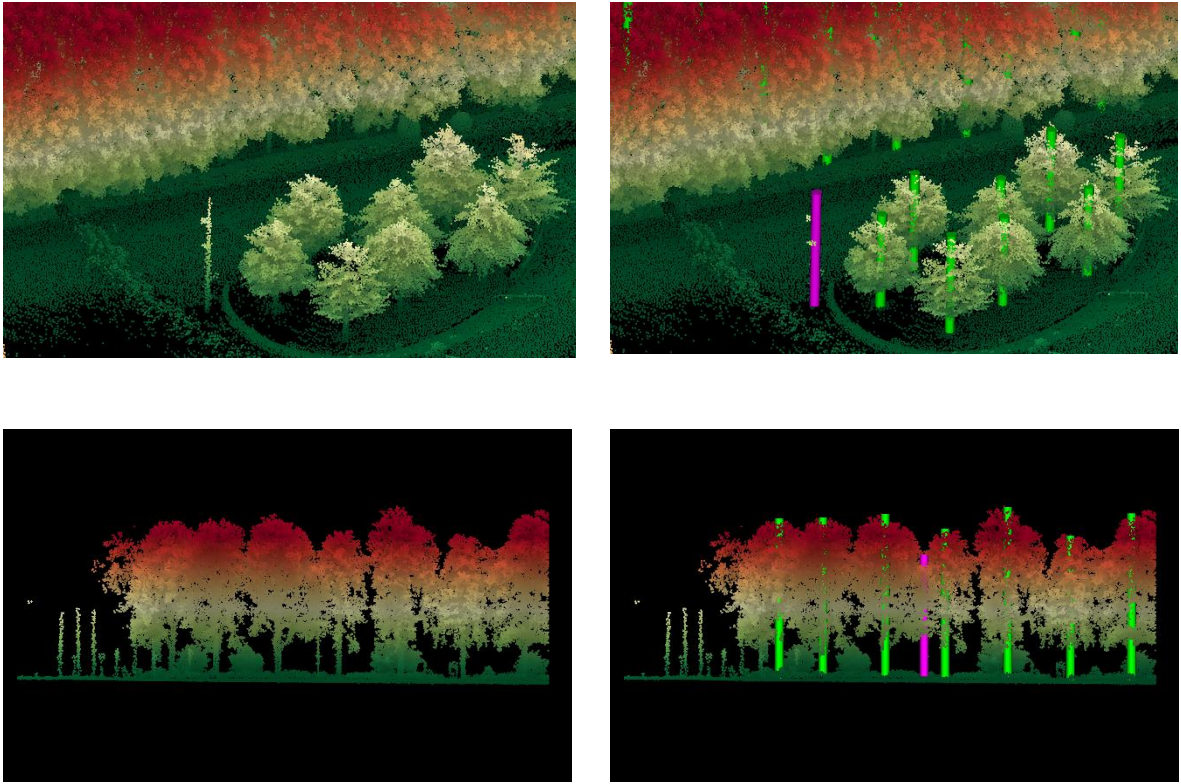

Figure 7: Exemplary point cloud profiles (left column) along with the integrated crowd acquisitions (right) classified as TP (green) and FP (pink) rooting from both Scene I (top) and Scene II (bottom), which are caused by crowdworkers confusing lamp posts and power line posts with tree stems.

Apart from these minor misclassifications, we would like to stress that our proposed methodology does not only yield high quality values as listed in Table 2, but is also time- and cost-efficient. For all scenes, respective campaigns ran for approx. 24 hours, meaning that an accurate map can be ready just the day after the collection of the 3D point cloud. With respect to cost-efficiency, on average, we payed $0.35 for each TP stem with overall costs depending both on the extent of the area of interest as well as the complexity of the scene, i.e., the more dense the vegetation, the narrower we should choose the tiles, which will eventually yield to higher costs per ha. Hence, for Scene IV, in which we are faced with a dense forest area, the cost is the highest per ha (but individual tree annotations are cheapest, cf. Table 2). Nevertheless, we believe that especially such densely vegetated forest areas are a suitable application area for our approach, since, such scenes are often not charted and the production of tree maps would require an enormous effort. Moreover, such areas are easy for crowdworkers to work on. Typically, beneath-crown vegetation is rather limited due to lack of light passing through dense tree crown layers.

In presence of dense beneath-crown vegetation, on the other hand, our approach is rather limited and is likely to yield unsatisfactory results – at least with respect to completeness. An example vividly illustrating this issue can be found in Figure 8. Although this is an excerpt of the tree alley scene (Scene I), some regions are characterized by dense hedges planted along such a tree alley. Due to this dense vegetation, hardly any tree stems are clearly visible. Thus, the only hint for presence of trees are distinct tree crowns, which, however, are not sufficient for crowdworkers to estimate correct stem positions. Additionally, the majority of workers make use of the "I see no stems"-button (cf. Figure 2) and are not to blame for doing so. Please note that the correct tree annotations in Figure 8 are limited to isolated trees standing in front of the described tree and hedge alley.

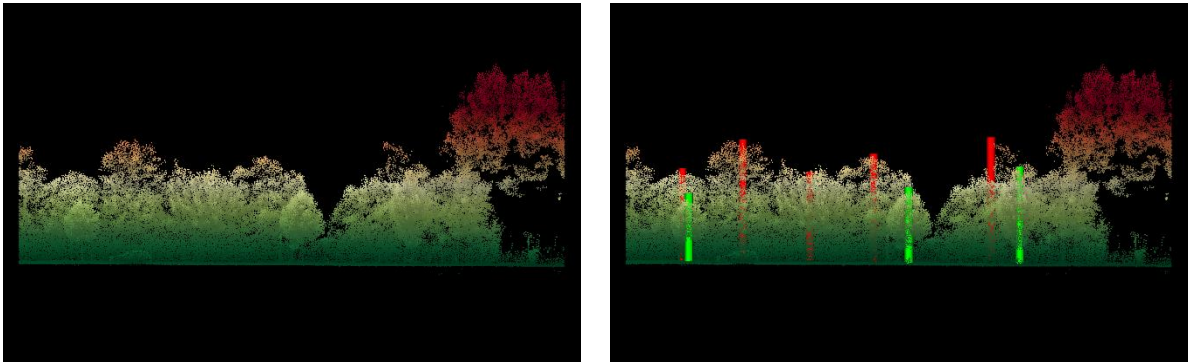

Figure 8: An exemplary point cloud profile (left) along with the crowd annotations (right) classified as TP (green) and FN (red). A high number of FNs occurs due to extensive beneath-tree-crown vegetation.

# 6 Conclusion

Within this paper, we have presented a crowd-based approach to generate tree maps in an accurate and timely manner. To achieve this, we recommend a self-sustaining device for data capturing like a handheld SLAM-based system, so that arbitrary scenes can be captured quickly. However, our approach is equally well suited for more sophisticated data sources that might be carried by airborne platforms. After pre-processing, that can be fully automatized, we leverage human processing power, i.e., we employ paid crowdworkers as human processing units. Those can be flexibly hired by means of respective crowdsourcing platforms such as *microWorkers*.

This requires having a web tool that is suited for crowdsourced data acquisition, i.e., we need to set-up a graphical user interface that allows workers without any experience with 3D data to do the required annotations. Developing this tool is demanding and time-consuming and was conducted in an engineering-like trial and error process.

Since we typically face a high degree of heterogeneity in quality of data collected by crowdowrkers, we developed methods to improve quality both during and after collection. For this purpose, we have realized an implementation of the Wisdom of the Crowds principle and developed a DBSCAN-based integration routine, which in turn can run fully automatically. Hence, apart from the initial data capturing step, all remaining steps can run without any expert involvement.

We see the main application of our approach in the generation of tree cadastres, since trees are often not mapped and the generation of corresponding maps is very expensive, if the data collection has to be done by experts.

# Conflict of Interest

The authors have no conflict of interest to declare that are relevant to this article.